\documentclass{ws-ijgmmp} 

\begin{document}

\markboth{Petr V. Tretyakov}
{Dynamical Stability of Minkowski Space in Higher Order Gravity} 

\title{DYNAMICAL STABILITY OF MINKOWSKI SPACE IN HIGHER ORDER GRAVITY}

\author{Petr V. Tretyakov}

\address{Bogoliubov Laboratory of Theoretical Physics, Joint Institute for Nuclear Research\\ 
Joliot-Curie 6, 141980 Dubna, Moscow region, Russia\\
\email{tpv@theor.jinr.ru} }

\maketitle
\begin{history}
\received{(Day Month Year)}
\revised{(Day Month Year)}
\end{history}

\begin{abstract}
We discuss the Minkowski stability problem in modified gravity by using dynamical system approach. The method to investigate dynamical stability of Minkowski space was proposed. This method was applied for some modified gravity theories, such as $f(R)$ gravity, $f(R)+\alpha R\Box R$ gravity and scalar-tensor gravity models with non-minimal kinetic coupling. It was shown that in the case of $f(R)$ gravity Minkowski solution asymptotically stable in ghost-free ($f'>0$) and tachyon-free ($f''>0$) theories in expanding Universe with respect to isotropic and basic anisotropic perturbations. In the case of higher order gravity with $\alpha R\Box R$ correction conditions of Minkowski stability with respect to isotropic perturbations significantly different: $f'(0)<0$, $f''(0)<0$ and $3f'(0)+f''(0)^2/\alpha>0$. And in the case of scalar-tensor gravity with non-minimal kinetic coupling Minkowski solution asymptotically stable in expanding Universe with respect to isotropic perturbations of metric. Moreover the developed method may be used for finding additional restrictions on parameters of different modified gravity theories.
\end{abstract}


\section{Introduction}

The unsolved problem of dark energy \cite{Riess,Perlmutter} generate a number of so-called modified gravity theories \cite{CFPS,Odintsov7}. One of the simplest from this modifications of gravity is $f(R)$ gravity \cite{Odintsov1,Tsujikawa,Odintsov2}, where scalar curvature $R$ in Einstein-Hilbert action is replaced with some function $f(R)$, so the action take the form

\begin{equation}
S=\int d^4x \sqrt{-g}f(R)+S_m.
\label{1.1}
\end{equation}
It is well known that equation of motion for this theory reads:
\begin{equation}
-\frac{1}{2}fg_{ik}+f'R_{ik}-\nabla_i\nabla_k f'+g_{ik}\Box f'=\kappa^2 T_{ik}.
\label{1.2}
\end{equation}
This equation contain higher derivatives with respect to metric up to the forth instead of the second one in General Relativity, and this fact may be the reason of different instabilities of classical solutions. There are two most general restrictions, which may be found by the different ways: $f'>0$ to guarantee that graviton is not a ghost (or to avoid antigravity on the classical level) and $f''>0$ to guarantee that particle associated with a new degree of freedom and named scalaron \cite{Starobinsky1} is not tachyon. There are may be different additional restrictions for $f(R)$ gravity also, which associated with another solutions, for instance Jeans instability \cite{CLMFO} or de Sitter stability condition \cite{AGPT}. Also such kind of theories may lead to the multiple de Sitter solution, which allow to describe inflation and late time acceleration within the unified approach \cite{Odintsov4,Odintsov5,Odintsov6}.
In this sense Minkowski solution takes a special place: from the one hand its stability is very important for theory by the obvious reasons, from another hand the investigation of this stability is very hard task by the usual dynamical system approach (which working good in de Sitter case), because of vanish eigenvalues. In this case we need to investigate central manifold from mathematical point of view and this task is much more 	
time-consuming. We can see the good illustration of troubles on this way in \cite{Miritzis}, where Minkowski stability was investigate in the simplest case of quadratic $f(R)$ gravity \cite{Starobinsky1}, which is still under consideration \cite{ADR,OGPR}. Some another considerations about Minkowski stability was discussed in \cite{Faraoni1}. In this paper we study stability of Minkowski solution by the dynamical system approach, but using some mathematical trick, which allow us obtain result without central manifold studying. The paper is organized as follows. In section 2 we develop our method and apply it to $f(R)$ gravity model. In section 3 we study Minkowski stability in one of the simplest case of higher order gravity with $R\Box R$ in the action. And in section 4 we turn to the scalar-tensor gravity model with non-minimal kinetic coupling. Some concluding remarks may be found in section 5.

\section{$f(R)$ gravity model}

First of all let us discuss FLRW metric
\begin{equation}
g_{ik}=diag(-1,a^2,a^2,a^2),
\label{1.3}
\end{equation}
with $\Lambda$-term as the simplest non-trivial matter
\begin{equation}
T_{ik}=diag(\Lambda,-\Lambda g_{11},-\Lambda g_{22},-\Lambda g_{33}).
\label{1.4}
\end{equation}
In this case full dynamical picture is described by the $00$-component of Eq. (\ref{1.2}) (for the sakes of simplicity we put $\kappa^2=1$):
\begin{equation}
\frac{1}{2}f-3(\dot H +H^2)f'+3H \dot f'=\Lambda,
\label{1.5}
\end{equation}
which can be rewritten as dynamical system of two variables $H$ and $R$:
\begin{equation}
\left \{
\begin{array}{l}
\dot H = \frac{1}{6}R-2H^2 \equiv F(H,R),\\
\\
\dot R= \frac{1}{3Hf''}\left (\Lambda -\frac{1}{2}f+\frac{1}{2}Rf'-3H^2f' \right ) \equiv G(H,R).
\end{array}
\right.
\label{1.6}
\end{equation}
It is clear that all equilibrium points of this system are de Sitter points (dS points) $(H_0,\,R_0)$ defined by the next relations
\begin{equation}
R_0=12H_0^2,\,\,S\equiv \frac{1}{2}f_0 - \frac{1}{4}R_0f'_0 -\Lambda=0.
\label{1.7}
\end{equation}
It's obvious that there are a lot of dS points in the most general case, but in this paper we will mainly discuss one from its, which is correspond to the Minkowski point $(H_0=0,\,R_0=0)$ at $\Lambda=0$. We denote it dS$_M$ (it is clear that for non-vanish values of $\Lambda$ this point is the nearest to the point $(H_0=0,\,R_0=0)$). This dS$_M$ point always exist even in the theories without intrinsic dS points. For example in quadratic gravity $f(R)=R+\alpha R^2$ there is no intrinsic dS points, but dS$_M$ point exist: it mean that there is one dS point (for non-vanish value $\Lambda$), which is tend to the Minkowski point as $\Lambda$ tend to zero ( for any $f(R)$-theory we have $R_0=12H_0^2\geqslant 0$ so $R_0\rightarrow +0$ as $\Lambda\rightarrow +0$). In this sense we may study Minkowski point as a limit point of dS$_M$ points set, and interpret $\Lambda$ as parameter of our theory\footnote{It is quite clear that nearest to zero dS point
   (in the case of non-vanish $\Lambda$) correspond to dS$_M$ point.}.

\subsection{Quadratic gravity}

Let us apply this idea to the simplest case of quadratic gravity to demonstrate how it works.
Thus we put
\begin{equation}
f(R)=R+\alpha R^2,
\label{1.8}
\end{equation}
and find from (\ref{1.7})
\begin{equation}
H_0=\sqrt{\frac{\Lambda}{3}},\,\,R_0=4\Lambda.
\label{1.9}
\end{equation}
Stability of this point is governed by the characteristic equation
\begin{equation}
\left |
\begin{array}{l}
(F_H)_0-\mu\,\,\,\,\,\,(F_R)_0\\
\\
(G_H)_0\,\,\,\,\,\,\,\,\,\,\,\,\,\,\,\,(G_R)_0-\mu
\end{array}
\right |=0,
\label{1.10}
\end{equation}
where (all this expressions are true for general case of $f(R)$-gravity)

\begin{equation}
F_H=-4H,
\label{1.11}
\end{equation}

\begin{equation}
F_R=\frac{1}{6},
\label{1.12}
\end{equation}

\begin{equation}
G_H=\frac{1}{6H^2f''}(f-Rf'-6H^2f'-2\Lambda),
\label{1.13}
\end{equation}

\begin{equation}
G_R=\frac{1}{6H(f'')^2}\left [ (R -6H^2)(f'')^2 + (6 H^2f' +f -Rf'  - 2 \Lambda)f'''  \right ],
\label{1.14}
\end{equation}
and index $0$ mean function's value at the studying stationary point, whereas $H$ or $R$ index mean partial derivative with respect to corresponding variable.

After the quite trivial calculations we find eigenvalues for quadratic gravity model (\ref{1.8})
\begin{equation}
\mu_{1,2}=\frac{1}{2}\left [ -\sqrt{3\Lambda} \pm \sqrt{3\Lambda - \frac{2}{3\alpha}(1+\alpha + 4\alpha\Lambda) }  \right ].
\label{1.15}
\end{equation}
We can see from this expression that from $\alpha >0$ and $\Lambda >0$ $\Rightarrow$ $\mathbf{Re} (\mu_{1,2})<0$. So we have $\mu_{1,2}\rightarrow -0\pm i\sqrt{\frac{2(\alpha +1 )}{3\alpha}}$ as $\Lambda\rightarrow +0$. It mean that any dS$_M$ point which is arbitrarily close to Minkowski point is stable and therefore we may say that Minkowski point is stable also and this result in good agreement with \cite{Miritzis}, where were find that Minkowski point is stable in quadratic $f(R)$ gravity in {\it expanding} universe (see below).

\subsection{General case of $f(R)$ gravity}

Now let us back to the general case of $f(R)$-gravity. Solution of Eq. (\ref{1.10}) reads
\begin{equation}
\mu_{1,2}=\frac{1}{2}\left [ \{(G_R)_0+(F_H)_0\} \pm \sqrt{ \{ (G_R)_0+(F_H)_0 \}^2 + 4(F_R)_0(G_H)_0 }  \right ],
\label{1.16}
\end{equation}
so stability conditions take the form
\begin{equation}
(G_H)_0<0,\,\,\,\, \{(G_R)_0+(F_H)_0\}<0.
\label{1.17}
\end{equation}
From another hand using (\ref{1.13}), (\ref{1.14}) and (\ref{1.7}) we find
\begin{equation}
(G_H)_0=-2\frac{f'_0}{f''_0},\,\,\,\, \{(G_R)_0+(F_H)_0\}=-3H_0.
\label{1.18}
\end{equation}
The most number of $f(R)$-gravity models apply $f'>0$ and $f''>0$ to avoid tachyon and ghost instability. Thus we have for exact dS solution $f'_0>0$, $f''_0>0$ and therefore any dS solution is stable with respect to homogeneous isotropic metric perturbations in expanding Universe ($H>0$).

Situation with Minkowski solution is not so trivial in the most general case. To discuss Minkowski stability problem let us back to the dS$_M$ point conception. We have next relation instead of (\ref{1.16})
\begin{equation}
\mu_{1,2}=\frac{1}{2}\left [ -3H_0 \pm 3H_0\sqrt{  1 - B }  \right ],
\label{1.19}
\end{equation}
where $B=\frac{4}{27}\frac{f'_0}{H_0^2f''_0}$ and $H_0=H_0(\Lambda)$, $f_0'=f_0'(\Lambda)$, $f_0''=f_0''(\Lambda)$ and $H_0\rightarrow +0$ as $\Lambda\rightarrow +0$. Since we imply $f'>0$ and $f''>0$ conditions we have a few different possibilities:
\begin{itemize}
\item case I\hspace{0.5cm}
$B\rightarrow +\infty$ as $\Lambda\rightarrow +0$. In this case we have $\mu_{1,2}= -\frac{3}{2}H_0\pm i\frac{3}{2}H_0\sqrt{B}$, so $Re(\mu_{1,2})\rightarrow -0$ and Minkowski solution is stable in the sense which was discussed above.

\item case Ia\hspace{0.5cm} $B\rightarrow B_0>1$ as $\Lambda\rightarrow +0$. In this case we have $\mu_{1,2}= -\frac{3}{2}H_0\pm i\frac{3}{2}H_0\sqrt{B-1}$, so $Re(\mu_{1,2})\rightarrow -0$ and Minkowski solution is stable also.

\item case II\hspace{0.5cm} $B\rightarrow B_0$, where $0<B_0<1$. In this case we have $\mu_1=-\frac{3}{4}H_0[B+O(B^2)]\rightarrow -0$ and $\mu_2=-\frac{3}{2}H_0[2-\frac{1}{2}B+O(B^2)]\rightarrow -0$ and $Im(\mu_{1,2})=0$, so Minkowski solution is stable.

\item case IIa\hspace{0.5cm} $B\rightarrow +0$. It's clear that in this case eigenvalues is similar to the previous one, so this case is a special case of the case II and Minkowski solution is stable.

\item case IIb\hspace{0.5cm} In the most trivial case $B\rightarrow 1$ both eigenvalues is equal to $-\frac{3}{2}H_0$, so Minkowski solution is stable also.

\end{itemize}

Thus we find that in any ghost-free and tachyon-free $f(R)$ gravity model Minkowski solution is stable with respect to homogeneous isotropic perturbations in expanding universe ($H>0$). Note that main number of models relate to the class I, for instance $R+\alpha R^n$, Hu-Sawicky \cite{HS}, Battye-Aplleby \cite{BA} or Starobinsky models \cite{Starobinsky3}.

\subsection{Anisotropic perturbations in $f(R)$ gravity}

Now let us turn to the more general case of homogeneous anisotropic perturbations.
For this task we need to discuss Bianchi I metric
\begin{equation}
g_{ik}=diag(-1,a^2,b^2,c^2),
\label{3}
\end{equation}
where functions $a$, $b$, $c$ are functions of time only. (It is well known that in GR all first type Bianchi metrics can be diagonalized and conserve its form due to Einstein equations. It is also true in $f(R)$-gravity, so expression (\ref{3}) is the most general form for Bianchi I metric in our case.) Also we introduce Hubble parameters $H_a=\dot a/a$, $H_b=\dot b/b$, $H_c=\dot c/c$. Only diagonal terms of equation (\ref{1.2}) is nontrivial and $(00)$ component now reads:
\begin{equation}
\frac{1}{2}f - (\dot H_a +H_a^2+\dot H_b +H_b^2+\dot H_c +H_c^2)f' + (H_a+H_b+H_c)\frac{\partial f'}{\partial t} =\Lambda.
\label{13}
\end{equation}

And for $(11)$ component we have:

\begin{equation}
-\frac{1}{2}f + (\dot H_a +H_a^2+H_aH_b+H_aH_c)f'+H_a\frac{\partial f'}{\partial t} - \frac{\partial^2 f'}{\partial t^2} -(H_a+H_b+H_c)\frac{\partial f'}{\partial t} =-\Lambda,
\label{14}
\end{equation}

for $(22)$:

\begin{equation}
-\frac{1}{2}f + (\dot H_b +H_b^2+H_aH_b+H_bH_c)f'+H_b\frac{\partial f'}{\partial t} - \frac{\partial^2 f'}{\partial t^2} -(H_a+H_b+H_c)\frac{\partial f'}{\partial t} =-\Lambda,
\label{15}
\end{equation}

for $(33)$:

\begin{equation}
-\frac{1}{2}f + (\dot H_c +H_c^2+H_cH_b+H_aH_c)f'+H_c\frac{\partial f'}{\partial t} - \frac{\partial^2 f'}{\partial t^2} -(H_a+H_b+H_c)\frac{\partial f'}{\partial t} =-\Lambda,
\label{16}
\end{equation}
So we have some strange situation: the highest derivative is contained in all three equations by the similar way: by the term $\ddot f'$ and therefore $\ddot R$ (note here that including of matter in r.h.s. do not change situation). So the system of differential equations is degenerated with respect to highest derivatives.

The interpretation of this fact may be the next. Actually the number of independent variables is {\it less} than 3. For illustration of this proposition let us try to transform system (\ref{13})-(\ref{16}). Introducing new variable $H\equiv H_a+H_b+H_c$ we find for expression in (\ref{13}):

$$
\dot H_a +H_a^2+\dot H_b +H_b^2+\dot H_c +H_c^2 = R - H^2 -\dot H,
$$

so the Eq. (\ref{13}) take the form:
\begin{equation}
\frac{1}{2}f - (R-\dot H - H^2)f' + H f''\dot R =0.
\label{17}
\end{equation}
From another hand summing Eqs. (\ref{14})$+$(\ref{15})+(\ref{16}) we find
\begin{equation}
-\frac{3}{2}f + (\dot H + H^2)f' - 2H f''\dot R -3f'''\dot R^2 - 3f''\ddot R =0.
\label{18}
\end{equation}
So we have actually system (\ref{17})-(\ref{18}) of two differential equations with two variables instead of system (\ref{14})-(\ref{16}) of three equations with three variables. Actually it do not mean that this is final result, because there may be further simplifications and this procedure is well known from the literature \cite{Starobinsky2}, where was shown how system (\ref{14})-(\ref{16}) can be transform to unique equation with one variable, but for our special task it is comfortable to use two variables: $H$ and $R$. Note one more time: this result is the general one for {\it any} $f(R)$-theories in Bianchi I ansatz (\ref{3})\footnote{Situation is totally equal to GR case, where in Bianchi I ansatz vacuum solution is described by the one parameter -- so called Kasner solution. This is true only for Bianchi I metric!}.

Let us consider this result in GR limit $f=R$, $\Lambda=0$. We have from Eqs. (\ref{17})-(\ref{18}):
\begin{equation}
\begin{array}{l}
\frac{1}{2}R - (R-\dot H - H^2)  =0,\\
-\frac{3}{2}R + (\dot H + H^2)  =0.
\end{array}
\label{19}
\end{equation}
We can see that situation is absolutely similar to the previous one: system is degenerated. It means that there is only one independent variable -- and this is true result, as we know from Kasner solution. System (\ref{19}) tell us that actually in this case there is only one equation $R=0$. Substituting this back to the system (\ref{19}) we find $\dot H +H^2=0$, which have solution $H=1/t$. From this relation we immediately reproduce one of the Kasner expressions $\sum p_i=1$.

Now let us study solutions of Eqs. (\ref{17})-(\ref{18}) using dynamical system approach. First of all we rewrite it as dynamical system:

\begin{equation}
\left \{
\begin{array}{l}

\dot H = -\frac{1}{f'}\left ( \frac{1}{2}f + H f'' D -\Lambda \right ) +R - H^2 \equiv F(H,R,D),\\
\\
\dot R=D,\\
\\
\dot D= \frac{1}{3f''} \left ( -2f - 3Hf''D -3f''' D^2 + R f' + 4\Lambda \right ) \equiv G(H,R,D).

\end{array}
\right.
\label{26}
\end{equation}

To investigate stability of some solution $(H_0,R_0,D_0)$ of the system (\ref{26}) we linearize it near this solution:
\begin{equation}
\left \{
\begin{array}{l}

\dot H = (F_H)_0H+(F_R)_0R+(F_D)_0D,\\
\\
\dot R=D,\\
\\
\dot D= (G_H)_0H+(G_R)_0R+(G_D)_0D,

\end{array}
\right.
\label{27}
\end{equation}
where $(F_H)_0$ denote the value of partial derivative of function $F$ with respect to $H$ at the point $(H_0,R_0,D_0)$ etc.
Characteristic equation for the linearized system (\ref{27}) is:
\begin{equation}
\left |
\begin{array}{l}
(F_H)_0-\mu \hspace{1cm} (F_R)_0 \hspace{1cm}  (F_D)_0
\\
\\
\,\,\,\,\,\,\,0  \hspace{2cm}  -\mu \hspace{1.5cm} 1
\\
\\
(G_H)_0 \hspace{1.5cm} (G_R)_0 \hspace{1cm}  (G_D)_0-\mu

\end{array}
\right | =0,
\label{28}
\end{equation}
which give us equation for eigenvalues $\mu$. It is easy to see that all equilibrium points are determined by the expression $D_0=0$. From another hand we have $(G_H)_0=-D_0=0$ for {\it any} equilibrium points of the system (\ref{26}), so actually we have instead of (\ref{28}):
\begin{equation}
((F_H)_0-\mu)\left |
\begin{array}{l}
-\mu \hspace{1.5cm} 1
\\
\\
(G_R)_0 \hspace{1cm}  (G_D)_0-\mu

\end{array}
\right | =0,
\label{28.1}
\end{equation}
which give us eigenvalues:
\begin{equation}
\mu_1=(F_H)_0,\,\,\,\, \mu_{2,3}=\frac{1}{2}\left ( (G_D)_0\pm\sqrt{(G_D)_0^2+4(G_R)_0} \right ),
\label{28.2}
\end{equation}
so stability conditions take the form
\begin{equation}
(F_H)_0=-2H_0<0,\,\,\,\, (G_D)_0=-H_0<0,\,\,\,\, (G_R)_0<0,
\label{28.3}
\end{equation}

It is easy to find that all equilibrium points are determined by the next expressions:
\begin{equation}
D_0=0,\,\,\,R_0=\frac{2f(R_0)-4\Lambda}{f'(R_0)},\,\,\,H_0^2=\frac{3}{4}R_0.
\label{35}
\end{equation}
This expressions are totally identical to (\ref{1.7}) with replacing $H\rightarrow 3H$, so we can see that all equilibrium points are some kind of dS points. The physical meaning of this points is not quite clear, since we have produce a some manipulations with initial variables, so first of all let us discuss it on the example of power-law function $f$, for which equations may be solved exactly:
\begin{equation}
f(R)=R+\alpha R^n.
\label{36}
\end{equation}
In the case $n=2$ there is only equilibrium point $(0,0,0)$. For $\alpha>0$ and $n>2$ there are three equilibrium points $(H_0,R_0,D_0)$: $(0,0,0)$, $(\pm\frac{\sqrt{3}}{2}[\alpha(n-2)]^{\frac{1}{2(1-n)}},[\alpha(n-2)]^{\frac{1}{(1-n)}},0)$. Let us discuss the physical meaning of this points.

 ${\bf (0,0,0)}$

Expression for $R$ may be rewritten as
\begin{equation}
R=2(\dot H + H^2 -H_aH_b -H_aH_c -H_bH_c),
\label{37}
\end{equation}
thus we have
\begin{equation}
\begin{array}{l}
H=0,\,\,\,\,\,\,\,\,\,\,\,\,\,\,\,\,\,\,\,\,H_a+H_b+H_c=0,\\
\,\,\,\,\,\,\,\,\,\,\,\,\,\,\,\,\,\,\,\,\,\,\,\,\,\,\Leftrightarrow\\
R=0,\,\,\,\,\,\,\,\,\,\,\,\,\,\,\,\,\,\,\,\,H_aH_b + H_aH_c + H_bH_c=0,
\end{array}
\label{38}
\end{equation}
which give us
\begin{equation}
H_b^2+H_bH_c+H_c^2=0.
\label{39}
\end{equation}
The only possibility to satisfy last expression (for non-complex values of Hubble parameters) is $H_a=H_b=H_c=0$, which is correspond to the Minkowski space. Note also that this point exist for any $f(R)$ gravity model with $f(0)=0$, $\Lambda =0$.

${\bf(\pm\frac{\sqrt{3}}{2} A, A^2, 0)}$

Here we introduce new notation $A=[\alpha(n-2)]^{\frac{1}{2(1-n)}}$. From (\ref{37}) we have:
\begin{equation}
H_aH_b +H_aH_c +H_bH_c=\frac{1}{4}A^2,
\label{40}
\end{equation}
and using definition of $H$
\begin{equation}
H_a = \pm\frac{\sqrt{3}}{2} A -H_b - H_c,
\label{41}
\end{equation}
we find the next expression
\begin{equation}
H_b^2 +H_b (H_c \mp\frac{\sqrt{3}}{2}A) +H_c^2 +\frac{1}{4}A^2 \mp\frac{\sqrt{3}}{2}AH_c =0,
\label{42}
\end{equation}
solving this equation with respect to $H_b$ we find discriminant
\begin{equation}
-3H_c^2\pm\sqrt{3}AH_c-\frac{1}{4}A^2,
\label{43}
\end{equation}
which can not be positive, but vanish at the point $H_c=\pm\frac{\sqrt{3}}{6}A$. Thus we have
\begin{equation}
(+\frac{\sqrt{3}}{2} A, A^2, 0)\Leftrightarrow H_a=H_b=H_c=\frac{\sqrt{3}}{6} A,
\label{44}
\end{equation}
which correspond to the usual dS point in expanding universe, and
\begin{equation}
(-\frac{\sqrt{3}}{2} A, A^2, 0)\Leftrightarrow H_a=H_b=H_c=-\frac{\sqrt{3}}{6} A,
\label{45}
\end{equation}
which correspond to the dS point in collapsing universe.

Now let us discuss stability conditions (\ref{28.3}). It's clear that $\mu_1<0$ for any $f(R)$ model in expanding universe. Expression for $\mu_{2,3}$ may be rewritten as
\begin{equation}
\mu_{2,3}=-\frac{1}{2}H_0 \pm\frac{1}{2}H_0\sqrt{1-\tilde{B}} ,
\label{46}
\end{equation}
where $\tilde{B}=\frac{4}{3}\frac{R_0f''_0 - f'_0}{H_0^2f''_0}$ and index $0$ mean function's value at dS point. Thus we reproduce well known condition ($\tilde{B}>0$) for stability of dS point \cite{Starobinsky4,Tsujikawa1}. As for dS$_M$ point we have situation absolutely similar to the previous one (isotropic perturbations). So for stability of Minkowski space in expanding universe it is enough one of the next conditions in any possible combinations:
\begin{itemize}

\item $f'\rightarrow +\infty$ or $f'\rightarrow A>0$ or $f'\rightarrow +0$  as $\Lambda\rightarrow +0$

\item $f''\rightarrow +\infty$ or $f''\rightarrow B>0$ or $f''\rightarrow +0$  as $\Lambda\rightarrow +0$

\end{itemize}

Thus our main conclusion is: Minkowski space is asymptotically stable in any tachyon-free ($f''>0$) and ghost-free ($f'>0$) $f(R)$ gravity model in expanding universe with respect to isotropic and basic anisotropic (homogeneous) perturbations.

\section{$f(R)+\alpha R\Box R$ gravity model}

Now let us discuss possible influence of higher derivative terms on Minkowski stability problem. As the simplest example of such kind of theory we study action in the next form
\begin{equation}
S=\int d^4x \sqrt{-g}[f(R) +\alpha R\Box R]+S_m.
\label{2.1}
\end{equation}

This theory is more complicate then usual $f(R)$ gravity model, so we discuss the simplest case of isotropic perturbations only. We have next additional terms in the left hand side of Friedman Eq. (\ref{1.5}) for FLRW metric (\ref{1.3}) (for more details see \cite{Tretyakov})
\begin{equation}
+\alpha ( 2R\ddot R +36H^3\dot R -\dot R^2 -48H^2\ddot R -12H \dddot R  ),
\label{2.2}
\end{equation}
so instead of system (\ref{1.6}) we have now

\begin{equation}
\left \{
\begin{array}{l}
\dot H = \frac{1}{6}R-2H^2,\\
\\
\dot R=C,\\
\\
\dot C=D,\\
\\
\dot D= \frac{1}{12\alpha H}\left (A -\Lambda +2\alpha RD+36\alpha H^3C - \alpha C^2 -48\alpha H^2D \right ) \equiv M,
\end{array}
\right.
\label{2.3}
\end{equation}
where $A=A(H,R,C)=\frac{1}{2}f+(3H^2-\frac{1}{2}R)f' + 3Hf''C$ is the left hand side of Eq. (\ref{1.5}). First of all note, that there is no additional dS-point due to (\ref{2.2})-terms, but this terms may change the stability conditions for dS point (including dS$_M$) arising from $f(R)$-part. Nevertheless in such kind of theory we have $R_0\rightarrow +0$ as $\Lambda\rightarrow +0$) as in the previous one. The linearized equation, which governs the stability, at equilibrium point takes the form
\begin{equation}
\left |
\begin{array}{l}
-4H_0-\mu \,\,\,\,\,\,\,\,\, \frac{1}{6} \,\,\,\,\,\,\,\,\,\,\,\,\,\,\,\,\, 0 \,\,\,\,\,\,\,\,\,\,\,\, \,\,\,\,\,\,\,\,\,\,\,0\\
\\
\,\,\,\,\,\,\,\,0\,\,\,\,\,\,\,\,\,\,\,\,\,\,\,\,\,\,\, -\mu \,\,\,\,\,\,\,\,\,\,\,\,\,\,\, 1 \,\,\,\,\,\,\,\,\,\,\,\,\,\,\,\,\,\,\,\,\,\,\, 0\\
\\
\,\,\,\,\,\,\,\, 0 \,\,\,\,\,\,\,\,\,\,\,\,\,\,\,\,\,\,\, \,\,\,\, \,\, 0 \,\,\,\,\, \,\,\,\,\, -\mu  \,\,\,\,\,\,\,\,\,\,\,\,\,\,\,\,\,\,\,\,\,\,\, 1\\
\\
(M_H)_0 \,\,\,\,\, \,\,\,\,\,\,(M_R)_0  \,\,\,\,\, (M_C)_0  \,\,\,\,\,(M_D)_0-\mu
\end{array}
\right |=0,
\label{2.4}
\end{equation}
or
\begin{equation}
a_0\mu^4+a_1\mu^3+a_2\mu^2+a_3\mu+a_4=0,
\label{2.5}
\end{equation}
with
\begin{equation}
\begin{array}{l}
a_0=1,\,\,a_1=4H_0-(M_D)_0,\,\,a_2=-4H_0(M_D)_0-(M_C)_0,\\
a_3=-(M_R)_0-4H_0(M_C)_0,\,\,a_4=-4H_0(M_R)_0-(M_H)_0/6.
\end{array}
\label{2.6}
\end{equation}
Since the finding of general solution of (\ref{2.5}) is a hard task we use Routh-Hurwitz theorem \cite{Gantmacher} which tell us that all solutions of (\ref{2.5}) have a negative real parts (and therefore equilibrium point $0$ is stable) if and only if satisfy next relations:
\begin{equation}
T_0=a_0>0,
\label{2.7}
\end{equation}
\begin{equation}
T_1=a_1>0,
\label{2.8}
\end{equation}

\begin{equation}
T_2=
\left |
\begin{array}{l}
a_1\,\,\,\,\,\,\,\,1
\\
a_3\,\,\,\,\,\,\,\,a_2
\end{array}
\right |>0,
\label{2.9}
\end{equation}
\begin{equation}
T_3=
\left |
\begin{array}{l}
a_1\,\,\,\,\,\,\,\,1\,\,\,\,\,\,\,\,\,\,\,0
\\
a_3\,\,\,\,\,\,\,\,a_2\,\,\,\,\,\,\,\,a_1
\\
0\,\,\,\,\,\,\,\,\,\,\,a_4\,\,\,\,\,\,\,\,a_3
\end{array}
\right |>0,
\label{2.10}
\end{equation}
\begin{equation}
T_4=
\left |
\begin{array}{l}
a_1\,\,\,\,\,\,\,\,1\,\,\,\,\,\,\,\,\,\,\,0\,\,\,\,\,\,\,\,\,\,0
\\
a_3\,\,\,\,\,\,\,\,a_2\,\,\,\,\,\,\,\,a_1\,\,\,\,\,\,\,\,1
\\
0\,\,\,\,\,\,\,\,\,\,\,a_4\,\,\,\,\,\,\,\,a_3\,\,\,\,\,\,\,\,a_2
\\
0\,\,\,\,\,\,\,\,\,\,\,0\,\,\,\,\,\,\,\,\,\,\,0\,\,\,\,\,\,\,\,\,\,\,a_4
\end{array}
\right |=a_4T_3>0.
\label{2.11}
\end{equation}

Now let us calculate partial derivatives of $M$. By using (\ref{1.7}) relations we find

\begin{equation}
(M_H)_0=\frac{f_0'}{2\alpha},
\label{2.12}
\end{equation}
\begin{equation}
(M_R)_0=-\frac{H_0f_0''}{2\alpha},
\label{2.13}
\end{equation}
\begin{equation}
(M_C)_0=\frac{f_0''}{4\alpha}+3H_0^2,
\label{2.14}
\end{equation}
\begin{equation}
(M_D)_0=-2H_0,
\label{2.15}
\end{equation}
and therefor
\begin{equation}
\begin{array}{l}
a_1=6H_0,\,\,a_2=5H_0^2-\frac{f_0''}{4\alpha},
a_3=-12H_0^3-\frac{H_0f_0''}{2\alpha},\,\,a_4=\frac{2H_0^2f_0''}{\alpha}-\frac{f_0'}{12\alpha}.
\end{array}
\label{2.16}
\end{equation}

We can see that (\ref{2.7}) and (\ref{2.8}) are satisfied automatically in expanding Universe ($H_0>0$), (\ref{2.9}) give us
\begin{equation}
42H_0^2 - \frac{f_0''}{\alpha}>0,
\label{2.17}
\end{equation}
from (\ref{2.10}) we find
\begin{equation}
-504H_0^4-81\frac{H_0^2f_0''}{\alpha}+\frac{f_0''^2}{\alpha^2}+\frac{3f_0'}{\alpha}>0,
\label{2.18}
\end{equation}
and (\ref{2.11}) give us
\begin{equation}
a_4=\frac{2H_0^2f_0''}{\alpha}-\frac{f_0'}{12\alpha}>0.
\label{2.19}
\end{equation}
In principle by using this inequalities we may verify stability of any dS point for any shape of function $f(R)$, but mainly we are interested in dS$_M$ point. For this point as we already mentioned we have $H^2\rightarrow+0$ as $\Lambda\rightarrow+0$.
From another hand it is well known \cite{Ovrut} that only positive values of $\alpha$ give us a ghost free theory. Thus from (\ref{2.17}) we have $f''_0<0$, from (\ref{2.19}) we have $f'_0<0$ and from (\ref{2.18}) we find $3f_0'+f_0''^2/\alpha>0$ (note also that for negative $\alpha$ the last inequality is impossible). This three conditions guarantee us stability of Minkowski solution with respect to isotropic perturbations. We can see that taking into account higher derivative terms may significantly change stability conditions for $f(R)$ gravity.

\section{Scalar-tensor gravity model with non-minimal kinetic coupling}

Now let us try to apply developed technique to the scalar-tensor gravity model with non-minimal kinetic coupling \cite{Odintsov3,Capozziello}

\begin{equation}
S=\int d^4x \sqrt{-g}[R - \{g^{\mu\nu}+\kappa G^{\mu\nu}\}\phi_{,\mu}\phi_{,\nu}-2V(\phi) - 2\Lambda],
\label{3.1}
\end{equation}
where we incorporate $\Lambda$-term in the action. Equations of motion for FLRW metric (\ref{1.3}) take the form \cite{Sushkov}
\begin{equation}
3H^2-\frac{1}{2}\dot\phi^2+\frac{9}{2}\kappa H^2\dot\phi^2=V+\Lambda,
\label{3.2}
\end{equation}
\begin{equation}
2\dot H + 3H^2 +\frac{1}{2}\dot\phi^2+\frac{1}{2}\kappa(2\dot H \dot\phi^2 + 3H^2\dot\phi^2 + 4H\dot\phi\ddot\phi)=V+\Lambda,
\label{3.3}
\end{equation}
\begin{equation}
(\ddot\phi+3H\dot\phi)-3\kappa(H^2\ddot\phi + 2H\dot H\dot\phi + 3H^3\dot\phi)=-V',
\label{3.4}
\end{equation}
where Eq. (\ref{3.2}) is the first integral of (\ref{3.3}) and (\ref{3.4}). We can see that this theory have second order equations, it mean that we may exclude $\dot H$ from the system without loss of generality. Indeed, multiplying Eq. (\ref{3.3}) on $6\kappa H\dot\phi$, Eq. (\ref{3.4}) on $(2+\kappa\dot\phi^2)$, summing and resolving with respect to highest derivative term, we gain the next dynamical system
\begin{equation}
\begin{array}{l}
\dot\phi = \Phi,\\
\\
\dot \Phi= \frac{-3H\Phi(2+3\kappa \Phi^2 - 6\kappa H^2 - 9\kappa^2 H^2 \Phi^2) - (2+\kappa \Phi^2)V'}{2+\kappa \Phi^2 -6\kappa H^2 + 9 \kappa^2 H^2 \Phi^2} \equiv f(H,\phi, \Phi),
\end{array}
\label{3.5}
\end{equation}
where $H$ now is not dynamical variable but parameter depending on $\Lambda$ and combination $V+\Lambda$ was excluded by using (\ref{3.2}).
Equilibrium points of system (\ref{3.5}) are defined by the next relations
\begin{equation}
\dot\phi_0=0,\,\,\,\,\, \ddot\phi_0=0,\,\,\,\,\, V'(\phi_0)=0,\,\,\,\,\,3H_0=\Lambda+V(\phi_0),
\label{3.6}
\end{equation}
where $V'(\phi_0)=0$ is the consequence of (\ref{3.4}). Since we are interested in Minkowski solution ($H_0=0$), we need to put also $V(\phi_0)=0$. Eigenvalues $\mu_{i}$ of system (\ref{3.5}) may be find from the next equation
\begin{equation}
\left |
\begin{array}{l}
-\mu\,\,\,\,\,\,\,\,\,\,\,\,\,\,\,\,\,\,1\\
\\
\,\,\,f_{\phi}\,\,\,\,\,\,\,\,\,f_{\Phi}-\mu
\end{array}
\right |=0,
\label{3.7}
\end{equation}
which have solution
\begin{equation}
\mu_{1,2}=\frac{1}{2}\left [ (f_{\Phi})_0 \pm(f_{\Phi})_0\sqrt{1+4\frac{(f_{\phi})_0}{(f_{\Phi})^2_0}} \right ],
\label{3.8}
\end{equation}
so the necessary and sufficient condition of equilibrium point's stability is
\begin{equation}
(f_{\phi})_0=\frac{-2V''(\phi_0)}{2-6\kappa H_0^2}<0,
\label{3.9}
\end{equation}
and
\begin{equation}
(f_{\Phi})_0=-3H_0<0.
\label{3.10}
\end{equation}
Thus we can see that Minkowski stability condition in expanding Universe ($H_0>0$) with respect to isotropic perturbations is $V''(\phi_0)>0$ (which is quite natural for any true vacuum solution), whereas stability of any nontrivial dS solution is governed by (\ref{3.9}) relation, where $H_0$ defined by $3H_0^2=\Lambda +V(\phi_0)$. Note also that Minkowski stability is not depend on sign of $\kappa$ parameter and this is the most unexpected result.

\section{Conclusion}

In this paper we propose a some universal asymptotic method for investigation stability of Minkowski solution in a wide class of modified gravity theories. The main idea quite simple: we introduce $\Lambda$-term as parameter and find eigenvalues of dS$_M$ point. After that we investigate limit of eigenvalues at $\Lambda\rightarrow +0$. This allows us to find Minkowski stability conditions. In some cases our method may be much more simple than any another one. So we hope it will be useful for a number researchers working in this field. Also we have applied our method to the some modified gravity theories and have found new original results (at least in Sec. 3). So we may conclude, that parameters of some theory, which lead to the instability of Minkowski space, are bad and must be excluded from further investigations. But we need keep in mind the next very important consideration, which have rather philosophical nature. Any our attempts to understand realistic world's picture are based on the our knowledge and we are forced to use our mathematical methodology even in those cases, where it may be not applicable. So we can say only that such kind of theories must be excluded only from the mathematical point of view, because we just don't know what the realistic picture is. There are may be some additional effects (even not discovered yet) like some quantum corrections or something else, which can allow realizing of the theories rejected earlier. The very good demonstration of this fact we can see in the paper: in pure $f(R)$ gravity we need $f'>0$ and $f''>0$, whereas taking into account higher derivative terms conversely give us $f'<0$ and $f''<0$ restrictions. In this sense all found in our paper areas of stability must be interpreted as necessary but not enough conditions of stability, which may be changed by additional more complicate effects.


\section{Acknowledgments}

This work was  supported by the RFBR grant 14-02-00894 A.

\end{document}